# Microwave properties of Yttrium Vanadate at cryogenic temperatures

Mohan V. Jacob[1)], Janina E. Mazierska[1)], Jerzy Krupka[1)], Dimitri O. Ledenyov[1)], Seiichi Takeuchi[2)]

1) Electrical and Computer Engineering Department, School of Engineering, James Cook University, Townsville, QLD4811, Australia.
2) Tokyo Denki University, 2-2, Kanda-Nishiki-Cho, Chiyoda-ku, Tokyo, Japan.

*Yttrium Vanadate* (*YVO_4*) is a birefringent crystal material used in optical isolators and circulators with potentials for application in cryogenic microwave devices. As microwave properties of the *YVO_4* are not known, we measured the complex permittivity at the frequency of *25 GHz*, using the *Hakki-Coleman* dielectric resonator technique in the temperature range from *13 K* to *80 K*. The real part of relative permittivity $\varepsilon_r$ turned out to be similar to that of *Sapphire* - one of popular dielectric materials, used at microwave frequencies. The measured loss tangent *tang $\delta$* of the *YVO_4* was of the order of *$10^{-6}$* at cryogenic temperatures. As *Yttrium Vanadate* (*YVO_4*) is easy to synthesis and machine, it may replace the expensive *Sapphire* in some microwave applications.

## Introduction

*Yttrium Vanadate* (*YVO_4*) is a birefringent crystal grown by the *Czochralski* method and is characterized by good mechanical and physical properties [1-3]. *YVO_4* is considered very suitable for optical polarizing components due to its wide transparency range and large birefringence. The *YVO_4* crystals are mainly used in optical components such as fiber optical isolators and circulators, beam displaces and other polarizing optical systems. The *Neodymium-doped Yttrium Vanadate* (*Nd:YVO_4*) is one of the most promising commercially available diode-pumped solid state laser materials for optical communication. The *YVO_4* has better thermal stability and physical and mechanical properties than $CaCO_3$, more than three times larger birefringence than $LiNbO_3$, and lower hardness than rutile $TiO_2$, what greatly reduces the cost of fabrication. Also, the *YVO_4* crystal growth and fabrication are easy than similar birefringent crystals.

Even though the *YVO_4* is well characterized at optical frequencies, there is hardly any data on microwave properties of this material. In this paper, we present, for the first time, the results of precise measurements of the permittivity and loss tangent of the *YVO_4* crystals fabricated by [3] at the frequency of *25 GHz* and the cryogenic temperatures from *13 K* to *80 K*. We have used a superconducting dielectric technique combined with the multi-frequency *Transmission Mode Q-Factor* (*TMFQ*) method [4, 5] for data processing to ensure the high accuracy of calculated values of $\varepsilon_r$ and tan $\delta$.

## Dielectric resonator measurement method

The dielectric resonator technique can be used for the microwave characterization of dielectric materials [6-10]. The permittivity and loss tangent of the dielectric material under test can be calculated from the unloaded *Q*-factor and the resonance frequency of the resonator. In our work, the *Hakki-Coleman* dielectric resonator, containing the *Yttrium Vanadate* (*YVO_4*) under test, schematically shown in Fig. 1, was used as the resonating structure. The resonator consisted of the cupper cavity with diameter of *9.5 mm* and height *3 mm* with *High Temperature Superconducting* (*HTS*) endplates to increase the sensitivity and reduce the uncertainty in the loss tangent measurements. The *YVO_4* sample was machined into the cylindrical shape sample with the aspect ratio (the diameter of height) equal to *1.61*, with the height of *3.09 mm* and the diameter of *4.99 mm*. The sample was oriented and machined in such a way that its *z*-axis was parallel to the crystal optical axis with accuracy better than *0.5 °*.

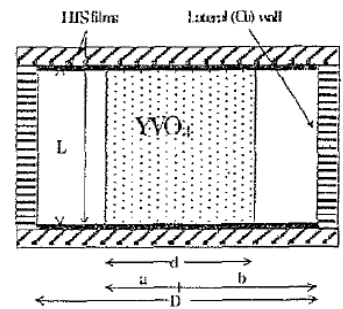

*Fig. 1. Hakki-Coleman dielectric resonator.*

Measurements were carried out, using the $TE_{011}$ mode of the resonator for which the electric field is perpendicular to z-axis, and hence the component of the complex permittivity of $YVO_4$ perpendicular to the optical axis was determined. The real part of relative permittivity $\varepsilon_r$ was determined from measurements of the resonance frequency as the first root of the following transcendental equation [11], using the software *SUP12* [12]

$$k_{\rho 1} J_0\left(k_{\rho 1} b\right) F_1(b) + k_{\rho 2} J_1\left(k_{\rho 1} b\right) F_0(b) = 0 \quad (1)$$

The loss tangent ($tan\ \delta$) of the $YVO_4$ was computed from the measured $Q_0$-factor of the resonator on the basis of the well known loss equation [11], namely

$$\tan \delta = \frac{1}{\rho_e}\left[\frac{1}{Q_0} - \frac{R_{SS}}{A_S} - \frac{R_{SM}}{A_M}\right] \quad (2)$$

where $Q_0$ is the unloaded Q-factor of the resonant structure with the sample, $R_{SS}$ and $R_{SM}$ are the surface resistances of the superconducting and the metallic parts of the cavity respectively, $A_S$ and $A_M$ are the geometric factors of the superconducting and the metallic parts of the cavity respectively, $\rho_e$ is the electric energy filling factor.

The geometric factors $A_S$, $A_M$ and $\rho_e$, to be used in eq. (2), were computed, using the incremental frequency rules [11] and calculated values for the $YVO_4$ and *Sapphire* (used for the measurements of $R_{SS}$ and $R_{SM}$ are given in Tab. 1.

| Dielectric | Yttrium Vanadate ($YVO_4$) | Sapphire |
|---|---|---|
| $f_{res}$ | 24.4 *GHz* | 24.65 *GHz* |
| $A_S$ | 19593 | 22319 |
| $A_M$ | 291.6 | 280.6 |
| $\rho_e$ | 0.97 | 0.97 |

Tab. 1. Calculated geometric factor of the dielectric resonator with *Yttrium Vanadate* ($YVO_4$) and *Sapphire* rods.

## Measurements of microwave properties of Yttrium Vanadate (YVO₄) crystals

The measurement system, we used for the microwave characterization of the $YVO_4$, is shown in Fig. 2. The system consisted of the Network Analyzer (*HP 8722C*), closed cycle refrigerator (*APD DE-204*), temperature controller (*LTC-10*) vacuum *Dewar*, a *PC*, and the *Hakki-Coleman* dielectric resonator in the transmission mode.

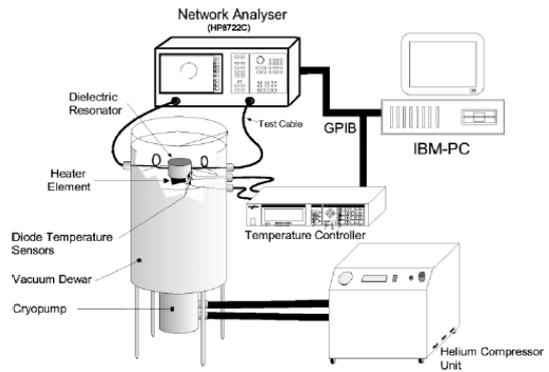

*Fig. 2. Experimental measurements set up.*

## Measurements of surface resistances of HTS thin film ($R_{SS}$) and copper walls ($R_{SM}$) of Hakki-Coleman cavity

The surface resistances of the *HTS* thin films $R_{SS}$ and the *Copper* walls $R_{SM}$, necessary for calculations of $tan\delta$ of $YVO_4$ sample, were measured in the same measurement system with the same dielectric resonator, but with the Sapphire rod instead of $YVO_4$. To obtain the precise values of surface resistances, we have measured the S-parameters ($S_{21}$, $S_{11}$ and $S_{22}$) around the resonance. The measured data sets were processed with the *Transmission Mode Q-Factor* (*TMFQ*) technique [4, 5] to obtain the loaded Q-factor ($Q_L$), coupling coefficients and unloaded Q-factor as mentioned in the Introduction. The *TMFQ* method accounts for the noise, delay due to the un-calibrated transmission lines and its frequency dependence, and crosstalk in measurement data, and hence provides the accurate values of the loaded Q-factor ($Q_L$) and the coupling coefficients $\beta_1$ and $\beta_2$. The unloaded Q-factor was subsequently calculated, using the exact equation

$$Q_0 = Q_L\left(1 + \beta_1 + \beta_2\right) \quad (3)$$

Assuming the loss tangent of the *Sapphire* rod as $10^{-7}$, the surface resistance of *Copper* $R_{SM}$ and superconductor $R_{SS}$ were calculated, using the eq. (2) [11]. The comprehensive information on the surface resistance $R_S$ measurements, using the *TMFQ* technique, is given in reference [5].

## Measurements of real part of relative permittivity $\varepsilon_r$ and loss tangent tan δ of Yttrium Vanadate (YVO₄) crystal

The *Hakki-Coleman* dielectric resonator with the HTS endplates, containing the $YVO_4$ sample, was cooled from the room temperature down to the temperature of *13 K* approximately, and the resonant frequency of *24.4GHz* was obtained. The $S_{21}$, $S_{11}$ and $S_{22}$ parameters data sets around the resonance were measured as a function of increasing temperature from *13 K* to *81 K*, and the $Q_0$-factor and $f_{res}$ were calculated, using the *TMFQ* technique and the eq. (3). The real part of relative permittivity $\varepsilon_r$ of the *Yttrium Vanadate* (*YVO$_4$*) sample was calculated from the measured resonant frequency, using the eq. (1) and taking the thermal expansion phenomenon into the consideration. The values of expansion coefficients ($\alpha$) of the *YVO$_4$* for the cryogenic temperatures were not available, and hence we assumed the thermal expansion coefficients as for another optical material *CaF$_2$* [14]. At the room temperature, the expansion coefficient ($\alpha$) for the *CaF$_2$* is around *17x10$^{-6}$/K*, while it is *8.5x10$^{-6}$/K* for the *YVO$_4$*. Therefore, our assumption of the expansion coefficient ($\alpha$) is most probably higher than the real values.

The measured real part of relative permittivity of *YVO$_4$* as a function of temperature $\varepsilon_r(T)$ at frequency of *24.4 GHz* in the temperatures range from *13 K* to *80 K* is shown in Fig. 3. The "circles" represent the values of the $\varepsilon_r$, computed assuming the thermal expansion coefficient of the *CaF$_2$*, and the "squares" represent the permittivity with the thermal expansion neglected. The difference between the values of permittivity with and without thermal expansion considered, is approximately *1 %*. The $\varepsilon_r$ exhibited the magnitude of approximately *9.3* and increased with the temperature by approximately *0.1 %*; from *9.318* to *9.326*.

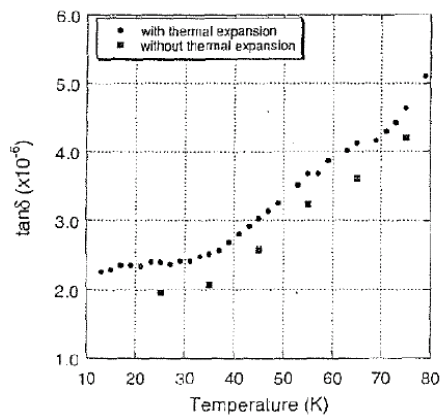

Fig. 4. Loss tangent of YVO$_4$ as a function of temperature $\tan\delta(T)$ at the frequency of 24.4 GHz.

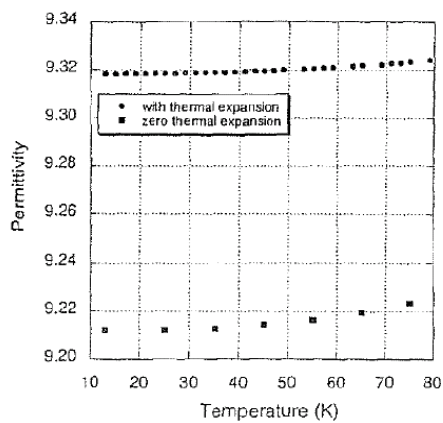

Fig. 3. Real part of relative permittivity of YVO$_4$ as a function of temperature $\varepsilon_r(T)$ at frequency of 24.4 GHz.

The Fig. 4 shows the measured temperature dependence of loss tangent of *YVO$_4$* at frequency of *24.4GHz*, calculated using the eq. (2) from the measured unloaded *Q*-factor. The loss tangent exhibits an increase of *125 %* in the temperatures range from *13 K* to *80 K*; the measured *tan $\delta$* of the *YVO$_4$* were *2.265x10$^{-6}$* and *5.1x10$^{-6}$* respectively. Using a linear scaling, the calculated loss tangent of the *YVO$_4$* at the temperature of *13 K* and the frequency of *10 GHz* is *9.2x10$^{-7}$* only.

## Conclusion

The complex permittivity of the *Yttrium Vanadate* (*YVO$_4$*) birefringent crystal has been precisely measured at the frequency of *24.4 GHz* at the cryogenic temperatures, using the *Hakki-Coleman* dielectric resonator with superconducting endplates. Our measurements accounted for the noise, crosstalk and uncompensated cables and adaptors, and were based on the accurate equations for the unloaded *Q*-factor. The *YVO$_4$* was found to exhibit the $\varepsilon_r$ varying from *9.31* to *9.32*, comparable to the permittivity of the *Sapphire*, and the *tan$\delta$* from *2.2x10$^{-6}$* to *5.1x10$^{-6}$* in the temperatures range from *13 K* to *80 K*. Our measurements have shown that the *YVO$_4$* is a very low loss material at the cryogenic temperatures. Therefore, apart from the optical applications, the *Yttrium Vanadate* (*YVO$_4$*) can be useful in the cryogenic microwave circuits as a replacement for the *Sapphire*.

## Acknowledgement

This work is supported by ARC-Large grant (*A00105170*) and ARC-Linkage International (*LX0242351*). The first author acknowledges the James Cook University Post Doctoral Fellowship and James Cook University MRG grant.

*E-mail: janina.mazierska@jcu.edu.au